\documentclass[twocolumn,times]{aastex62}
\usepackage{amsmath}
\usepackage{amssymb}
\usepackage{lineno}

\usepackage{bm}
\allowdisplaybreaks[4]

\def\be {\begin{equation}}
\def\ee {\end{equation}}

\def\tt {\{t\}}

\usepackage{color}
\usepackage{booktabs}
\shorttitle{Origin of the Moon}
\shortauthors{Wenshuai Liu}

\begin{document}

\title{\large{\textbf{Origin of the lunar isotopic crisis from solidification of a stratified lunar magma ocean}}}

\correspondingauthor{Wenshuai Liu}
\email{674602871@qq.com}

\author{Wenshuai Liu}
\affiliation{School of Physics, Henan Normal University, Xinxiang 453007, China}



\begin{abstract}
According to the giant impact theory, the Moon formed through accreting the debris disk produced by a collision between Theia and the proto-Earth. The giant impact theory can explain most of the properties of the Earth-Moon system, however, simulations with respect to giant impact between a planetary embryo and the growing proto-Earth show that more than 40 percent of the materials in the circum-terrestrial debris disk produced by the giant impact originates from the impactor. Thus, the giant impact theory has difficulty explaining the Moon's Earth-like isotopic compositions, which is referred to as the lunar isotopic crisis. With the assumption that Theia may have possessed an iron-rich mantle compared with proto-Earth's mantle, here we show that, after the formation of the stratified lunar magma ocean, solidification of the stratified lunar magma ocean would result that the upper solid layer is composed of proto-Earth's mantle and the lower solid layer is made of Theia's mantle, meaning that the Moon's Earth-like isotopic composition is a natural result of the giant impact. The theory proposed here may provide a way of explaining the lunar isotopic crisis.
\end{abstract}



\section{Introduction}

The giant impact theory \citep{1,2} for the lunar origin is thought to lead to the formation of the Moon through accreting material from a circum-terrestrial debris disk after the impact of a planet-sized object with the proto-Earth. Numerical simulations based on this theory have revealed that the circum-terrestrial debris disk, from which the Moon formed, comprises a large portion of materials from the impactor \citep{15,16,17}. Given that various bodies within the Solar System exhibit distinct compositional characteristics, the lunar composition, as inferred from the canonical giant impact model, would diverge notably from that of Earth. Conversely, comparative isotopic analyses of samples from both Earth and the Moon demonstrate a striking similarity in their isotopic signatures \citep{18,19,20,21,22,23}, leading to what is known as the lunar isotopic crisis. To account for the observed isotopic similarity between Earth and its Moon, materials forming the Moon should equilibrate with or be from the Earth's mantle after the impact \citep{24,22}, or the Moon originated from the giant impact with an impactor possessing isotopic characteristics identical to those of Earth \citep{26}. Nonetheless, these models have not successfully reconciled all geochemical observations, indicating a need for further investigation into the complex processes governing Earth-Moon formation and composition.

\cite{27} proposed a modified version of the giant impact theory that incorporates a rapidly rotating proto-Earth. In this model, a greater portion of material from proto-Earth could be delivered to the circum-terrestrial debris disk, which is predominantly composed of material from the proto-Earth. This disk is sufficiently massive to potentially form the Moon following the collision of a body slightly smaller than Mars with the fast-spinning proto-Earth. This hypothesis has the potential to explain the observed isotopic composition of the Moon, however, it results in an angular momentum that exceeds that of the contemporary Earth-Moon system. Other models characterized by high energy and high angular momentum, including Synestia \citep{28,29} and the collision of two bodies of comparable mass \citep{30}, similarly yield substantial angular momentum that must be reduced through a viable mechanism that remains in debate. The high-energy, high-angular-momentum model \citep{27} aims to establish a circum-terrestrial debris disk primarily composed of materials from the proto-Earth's mantle, thereby accounting for the Earth-like isotopic composition of the Moon. What happens if the materials in the mantle of Theia have an equation of state different from that of the materials in the mantle of proto-Earth? Such different equation of state may significantly influence the formation processes and resultant characteristics of the Moon.

Since smaller bodies, like Vesta and Mars, typically have a greater proportion of iron in their mantles compared with Earth \citep{27,31}, Theia may have possessed an iron-rich mantle compared with the proto-Earth's mantle \citep{32}. This suggests that, under the same pressure and temperature, the density of materials of Theia's mantle is greater than that of proto-Earth's mantle. Consequently, after giant impact, a moon forms within the circum-terrestrial molten disk, where the blobs of molten material originating from Theia's mantle would sink to the lower layer after they accrete onto the moon, while the blobs of molten material from proto-Earth's mantle would float in the upper layer. Then, a stratified lunar magma ocean formed inside the Moon.

In this work, we confirm that, as the stratified magma ocean cools, both layers would solidify with the resulting upper solid layer composed of proto-Earth's mantle and lower solid layer made of Theia's mantle. From this perspective, the Moon's Earth-like isotopic composition emerges as a natural consequence of the giant impact. The detailed demonstration is in Section 2 and Section 3, and the discussions are given in Section 4.

\section{Formation of the stratified lunar magma ocean}

After the giant impact, circum-terrestrial material ejected by giant impact is predominantly silicate melt with temperature of 2000K to 5000K exterior to the Roche limit \citep{37}. Due to the longer timescale of cooling of the disk to temperatures below the solidus compared with the timescale of formation of Moon through accretion in the outer disk \citep{36}, the Moon formed by accreting material orbiting outside the Roche limit would be in a hot and molten state after formation.

\begin{table}
\centering
\caption{Liquid compositions in proto-Earth and Theia}
\begin{tabular}{ccc}
\toprule
wt$\%$ oxides & proto-Earth  & Theia \\
\midrule
SiO$_2$ & 44.9 & 43.90 \\
Al$_2$O$_3$ & 4.44 & 3.15 \\
MnO & 0.13 & 0.0 \\
Cr$_2$O$_3$ & 0.38 & 0.0 \\
Fe$_2$O$_3$ & 0.116 & 0.0 \\
FeO & 7.926 & 18.8 \\
MgO & 37.71 & 31.66 \\
CaO & 3.54 & 2.50 \\
Na$_2$O & 0.36 & 0.0 \\
K$_2$O & 0.029 & 0.0 \\
P$_2$O$_5$ & 0.021 & 0.0 \\
H$_2$O & 0.45 & 0.0 \\
\bottomrule
\end{tabular}
\end{table}

We assume that Theia has a higher proportion of iron in its mantle than that of proto-Earth's mantle. Thus, during formation of Moon through accretion of circum-terrestrial molten disk after the giant impact, a stratified magma ocean forms due to the fact that molten materials originating from Theia's mantle sink to the lower layer after accreting onto the Moon while molten materials from proto-Earth's mantle float in the upper layer. This outcome can be verified by taking the detailed composition of Theia and proto-Earth. Here, the composition of Theia's mantle is same to that of \cite{39} adopted to Mars shown in Table 1 where the composition of proto-Earth's mantle, similar to that of \cite{38} except that TiO$_2$ and NiO are replaced by H$_2$O,is also given. With the composition of Theia's mantle and proto-Earth's mantle, the density of the blobs of molten material from Theia's mantle is about $2.78\mathrm{g/cm^3}$ generated by alphaMELTS when pressure is 0GPa and temperature is 2000$^\circ \mathrm{C}$. Similarly, the density of the blobs of molten material from proto-Earth's mantle is about $2.67\mathrm{g/cm^3}$ generated by alphaMELTS when pressure is 0GPa and temperature is 2000$^\circ \mathrm{C}$. In order to test the blobs of molten material from Theia's mantle cannot mix with the surrounding molten material from proto-Earth's mantle during sinking after accreting onto the surface of Moon, the timescale of mixing should be much longer than the timescale of sinking. This can be confirmed in details as follows.

The radius of the blobs of molten material from Theia's mantle is taken to be $\mathrm{r=1km}$. In the magma ocean made of molten material from proto-Earth's mantle, due to the large radius of the blobs of molten material from Theia's mantle, the drag force acting on the blobs of molten material from Theia's mantle is approximated as follows when the velocity of steady state is reached
\begin{equation}
F_d=\frac{1}{2}C_d\rho_E A v^2 \label{1}
\end{equation}
where $C_d$ is drag coefficient, $A=\pi r^2$ is the characteristic area, $\rho_E$ is the density of molten material from proto-Earth's mantle and $v$ is the steady state velocity.

The Moon's surface gravitational force acting on the blobs of molten material from Theia's mantle is
\begin{equation}
F_g=\frac{4}{3}\pi r^3 (\rho_T-\rho_E)g \label{2}
\end{equation}
where $\rho_T$ is the density of the blobs of molten material from Theia's mantle and $g$ is the Moon's surface gravitational acceleration.

With Eq. (\ref{1}) and Eq. (\ref{2}), the steady state velocity can be obtained and is given as
\begin{equation}
v= \sqrt{\frac{8r(\rho_T-\rho_E)g}{3C_d\rho_E}} \label{3}
\end{equation}

With the present value of the Moon's surface gravitational acceleration and $C_d=0.45$, we get $v\approx20\mathrm{m/s}$. The Reynolds Number $Re=\frac{\rho_E v r}{\mu}\approx5.3\times10^6\gg1$ when viscosity of the blobs of molten material from proto-Earth's mantle $\mu=10\mathrm{Pa\cdotp s}$. This means that Eq. (\ref{1}) is suitable to describe the dynamics of the sinking of the blobs of molten material from Theia's mantle in the magma ocean made of molten material from proto-Earth's mantle. Then the timescale of sinking of the blobs of molten material from Theia's mantle is $t=\frac{R}{v}\approx23$ hours where the radius of Moon $R$ is taken to be $R=1700km$.

The timescale of mixing can be obtained as $t=\frac{L^2}{D}$ where $D$ is the diffusion coefficient and $L$ is the diffusion distance. Usually, $D$ is order of $D=10^{-11}\mathrm{m^2/s}$ in magma, then we get $t\approx3000$ years if $L=1m$ is adopted. This means that the timescale of mixing is much longer than the timescale of sinking and that, during the sinking of the blobs of molten material from Theia's mantle, the mixing of blobs of molten material from Theia's mantle and that from proto-Earth's mantle is negligible. Then, as shown in Figure 1, a stratified lunar magma ocean would form inside the Moon after formation.

\begin{figure}
            \includegraphics[width=0.5\textwidth]{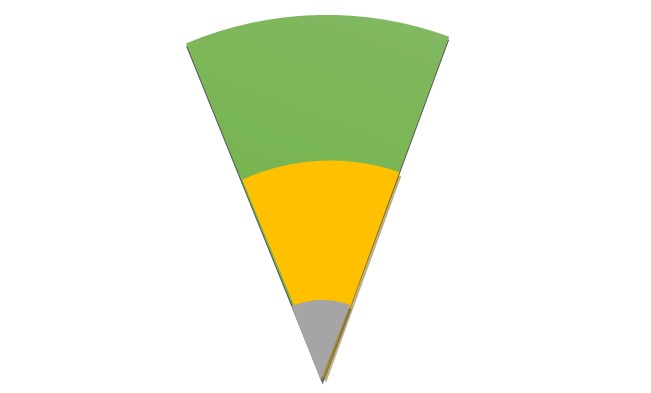}
\caption{A stratified magma ocean forms after molten materials originating from Theia's mantle sink to the lower layer (orange color) after accreting onto the moon while molten materials from proto-Earth's mantle float in the upper layer (green color). The gray region is the lunar core.}
\label{fig:figure1}
\end{figure}

\section{Solidification of the stratified lunar magma ocean}

Due to the larger compressibility of melts relative to solids under high pressure, olivine, orthopyroxene and clinopyroxene may have neutral buoyancy or even positive buoyancy during fractional crystallization in the deep region of the lunar magma ocean. Studies in \cite{34} show that the first olivine to crystallize in lunar magma ocean is neutrally buoyant at 3.8 GPa, corresponding to at about 1150km depth, and would accumulate there. Researches in \cite{35} demonstrate that, due to the high iron content in Martian mantle and the partitioning of iron into the liquids, the onset of negatively buoyant magmas occurs at lower pressures on Mars, near 3GPa. In lunar magma ocean, 3GPa corresponds to at about 700km depth, meaning that the iron-rich magma in the lower layer shown in Figure 1 would decrease the pressure for neutrally buoyant olivine and the corresponding depth.

Based on the presence of the stratified lunar magma ocean, Moon's Earth-like isotopic composition would be a natural results if both of the upper magma ocean composed of proto-Earth's mantle and the lower magma ocean made of Theia's mantle solidify without crystals floating from the lower magma ocean to the upper during fractional crystallization. In order to show whether the process of fractional crystallization could produce the isolation
of the two layers, we use code the pMELTS \citep{49} in alphaMELTS to simulate the fractional crystallization with liquid compositions shown in Table 1. The fractional crystallization of the two layers near the interface is simulated to test whether accumulation of crystal could emerge near the interface of the two layers. Isolation of the two layers through accumulation of crystal near the interface would further produce two solid layers with the upper layer composed of proto-Earth's mantle and the lower made of Theia's mantle after solidification of the stratified magma ocean. The pressure under which melts crystallize is taken to be at $\mathrm{P=1.5GPa, 2GPa, 2.5GPa}$ and $\mathrm{3GPa}$ and the liquidus temperature is taken as the starting temperature.

\begin{figure*}
     \begin{tabular}{cc}
            \includegraphics[width=0.5\textwidth]{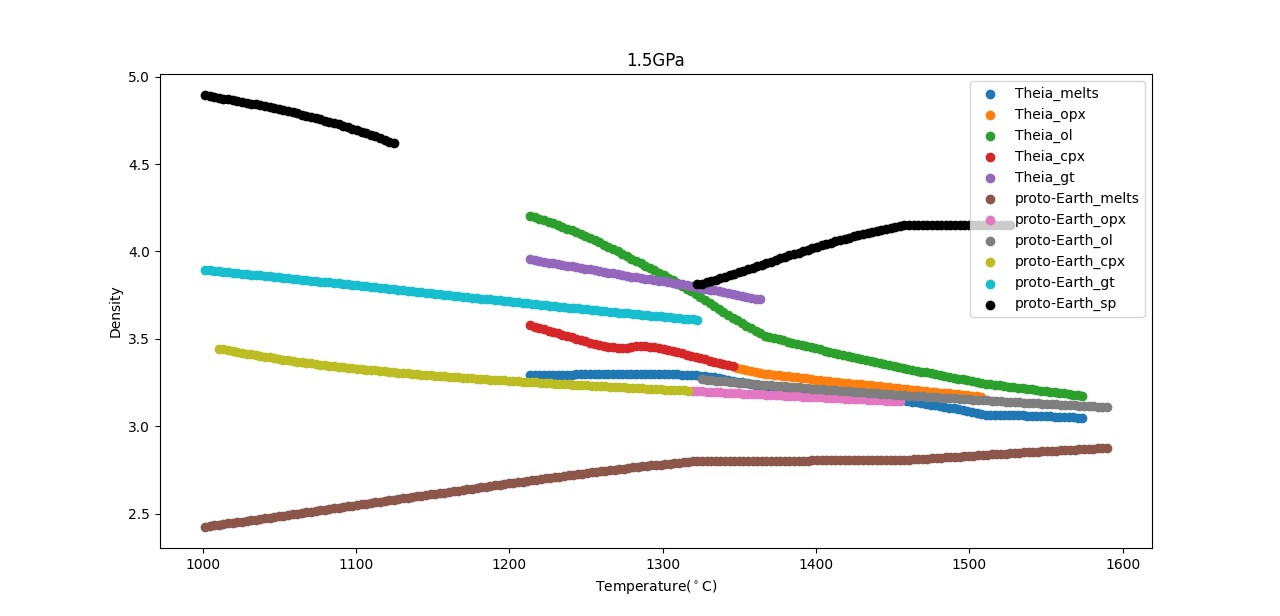}
            \includegraphics[width=0.5\textwidth]{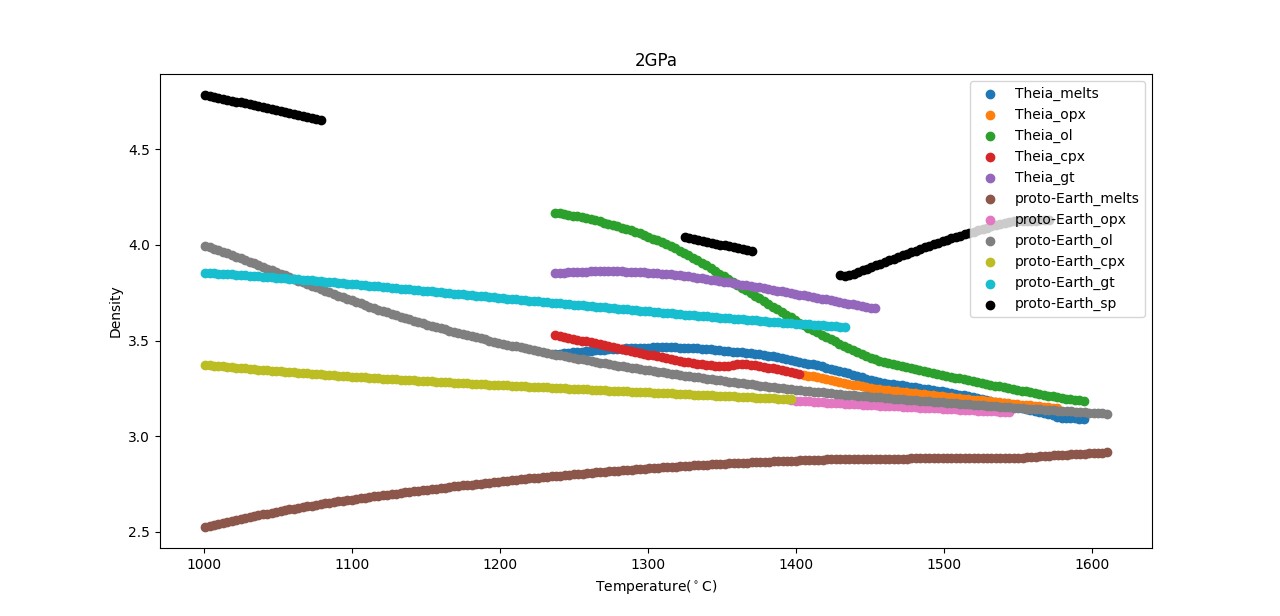}\\
            \includegraphics[width=0.5\textwidth]{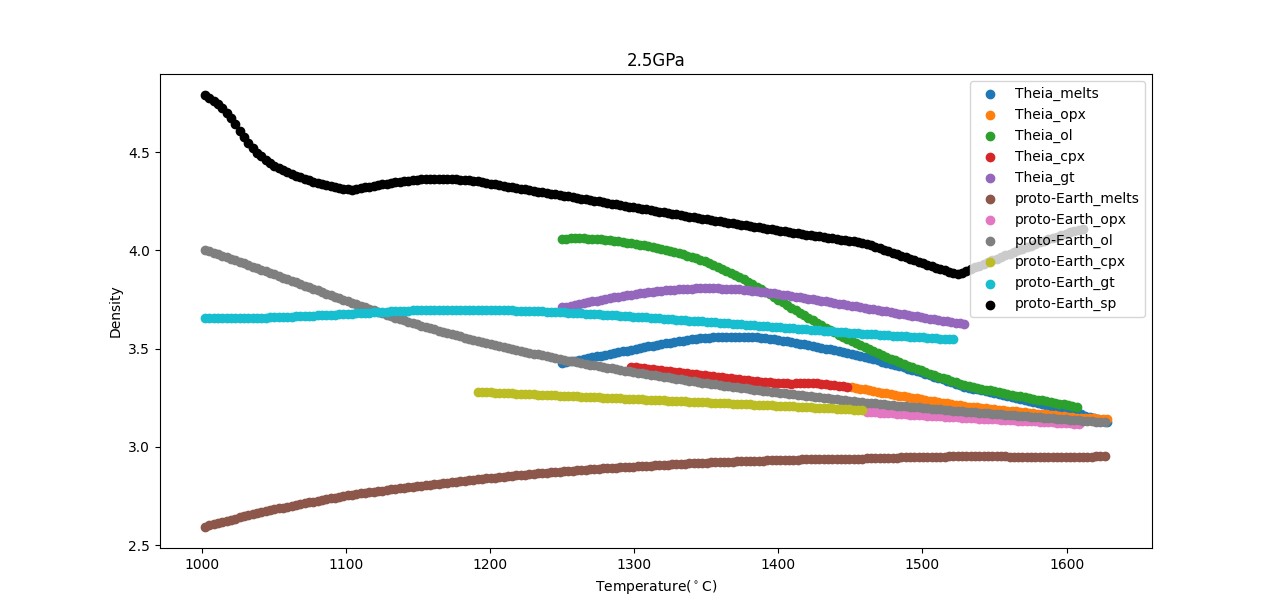}
            \includegraphics[width=0.5\textwidth]{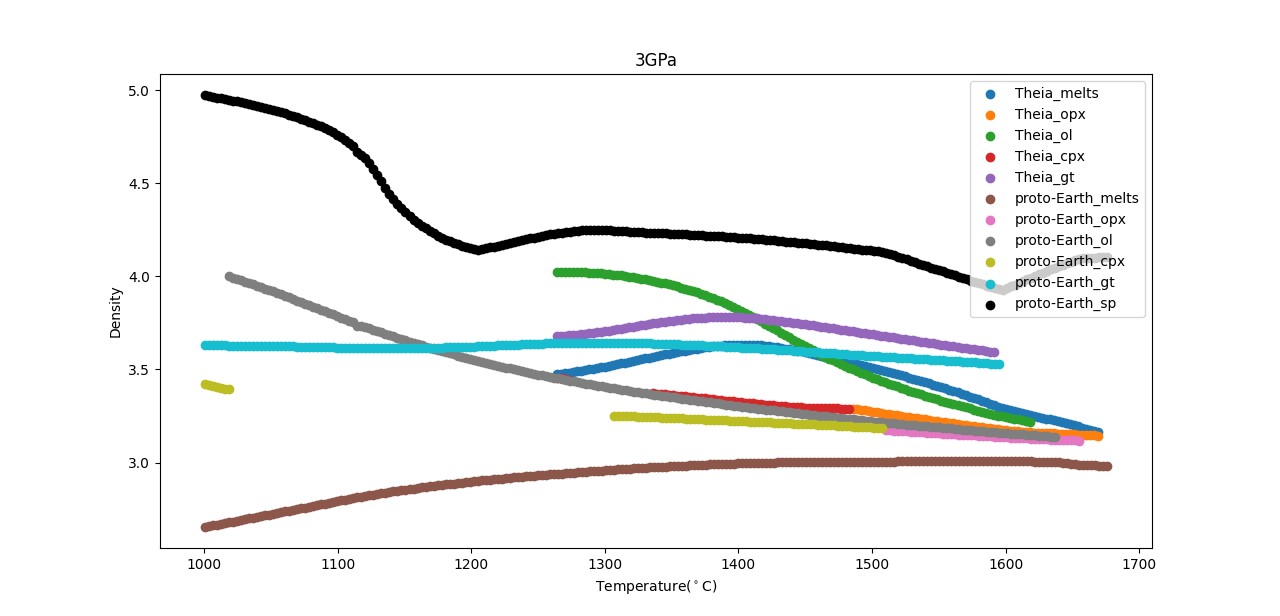}
            \end{tabular}
\caption{The density of the melts and different crystals produced by fractional crystallization during cooling of the magma ocean under different pressure, density is in unit of $\mathrm{g/cm^3}$. opx = orthopyroxene; ol = olivine; gt = garnet;
sp = spinel; cpx = clinopyroxene. Theia$\mathrm{\_}$ and proto-Earth$\mathrm{\_}$ represent the origin of the melts and different crystals.}
\label{fig:figure2}
\end{figure*}

Figure 2 shows the density of the melts and the crystals produced by fractional crystallization during cooling of the magma ocean under different pressure. Here, for simplicity, we take the results under pressure $\mathrm{2GPa}$ for detailed analysis. During fractional crystallization, the density of the orthopyroxene and clinopyroxene produced by the upper melts near the interface is always lower than the density of the lower melts, meaning that the orthopyroxene and clinopyroxene crystallizing from the upper melts would float on the lower melts. About $27\%$ of the upper melts near the interface in mass is in the form of the orthopyroxene and clinopyroxene floating on the lower melts after completion of the fractional crystallization of the upper melts. At about $\mathrm{1549^\circ C}$, the density of olivine crystallizing in the upper melts begins to be lower than the density of the lower melts and starts to float on the the lower melts and the density difference reverses at $\mathrm{1240^\circ C}$. From $\mathrm{1549^\circ C}$ to $\mathrm{1240^\circ C}$, about $16\%$ of the upper melts near the interface in mass is in the form of the olivine floating on the lower melts. When temperature continues to decrease, about $1\%$ of the upper melts near the interface in mass is in the form of the olivine with density larger than the density of the lower melts. Although the larger density of the $1\%$ in the form of the olivine relative to the density of the lower melts, such olivine will accumulate on the top of the olivine formed before. At about $\mathrm{1531^\circ C}$, the orthopyroxene produced by the lower melts near the interface begins to float on the lower melts. When temperature decreases to about $\mathrm{1285^\circ C}$, the density of clinopyroxene crystallizing in the lower melts begin to be larger that the lower melts. From $\mathrm{1531^\circ C}$ to $\mathrm{1258^\circ C}$, about $13\%$ of the lower melts near the interface in mass is in the form of the orthopyroxene and clinopyroxene floating on the lower melts.

The results obtained under $\mathrm{2GPa}$ is analyzed in details, the pressure $\mathrm{P=2GPa}$ is at a depth of about $\mathrm{420km}$ or at radius of about $\mathrm{1300km}$. The mass within such radius is about $45\%$ the mass of the Moon if uniform density of the Moon is assumed.

Based on the analysis above, Moon's Earth-like isotopic composition is a natural results after both of the upper magma ocean composed of proto-Earth's mantle and the lower magma ocean made of Theia's mantle solidify. The isolation is much more obvious if the interface is under $\mathrm{2.5GPa}$ and $\mathrm{3GPa}$. Thus, as shown in Figure 3, stratified solid layers would be formed.

\begin{figure}
            \includegraphics[width=0.5\textwidth]{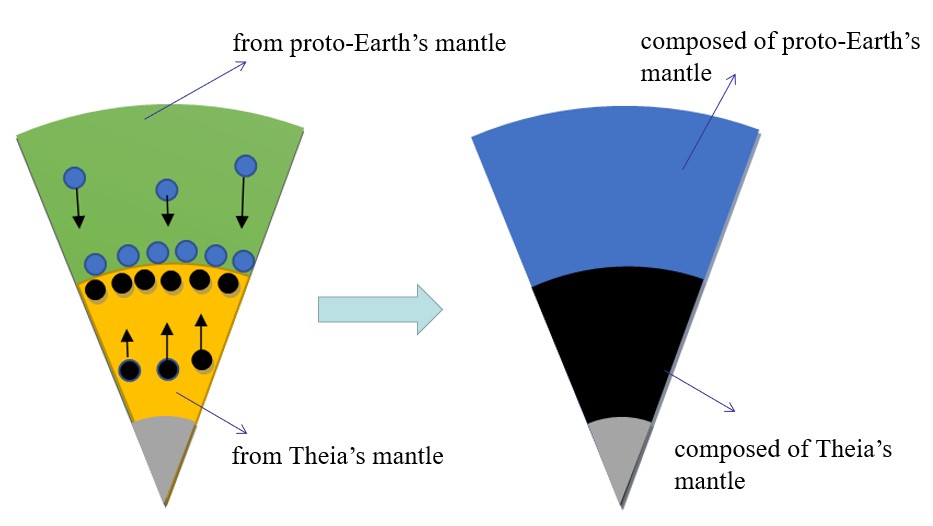}
\caption{Left: As magma ocean cools, the crystals from the upper layer (blue dots) accumulate at the boundary if the density of the crystals from the upper layer larger than that of the melt of the upper layer is lower than that of the magma ocean's lower layer. Accumulation of the crystals (black dots) from the lower layer at the boundary would enhance the isolation of the two layers if the density of crystals from the lower layer is lower than that of the melt of the lower layer but larger than that of the melts of the upper layer. Right: Solidification of the two layers where the blue layer is made of proto-Earth's mantle and the black layer is composed of Theia's mantle.}
\label{fig:figure1}
\end{figure}

\section{Discussions}
Lunar isotopic crisis is explained in this work. Under the assumption that Theia may have an iron-rich mantle compared with the proto-Earth's mantle, after giant impact, molten material originating from Theia's mantle would sink to the lower layer after accreting onto the moon while molten material from proto-Earth's mantle would float in the upper layer due to the fact that density of materials of Theia's mantle is larger than that of proto-Earth's mantle under the same pressure and temperature, indicating that the observed isotopic similarity between Earth and Moon may be a natural result of giant impact.

\section*{ACKNOWLEDGEMENTS}
We are grateful for the code alphaMELTS. This work is supported by Natural Science Foundation of Henan (252300420327).

\end{document}